\documentclass[sigconf, nonacm]{acmart}
\AtBeginDocument{%
  \providecommand\BibTeX{{%
    Bib\TeX}}}

\usepackage{enumitem}
\usepackage{siunitx} 
\usepackage{comment}

\usepackage{subfigure}
\usepackage{pifont}
\usepackage[linesnumbered,ruled,vlined]{algorithm2e}

\usepackage{multirow}
\usepackage{booktabs}
\usepackage{soul}
\usepackage{lipsum}
\usepackage{threeparttable}

\usepackage{algorithmic}
\usepackage{graphicx}
\usepackage{textcomp}
\usepackage{xcolor}
\def\BibTeX{{\rm B\kern-.05em{\sc i\kern-.025em b}\kern-.08em
    T\kern-.1667em\lower.7ex\hbox{E}\kern-.125emX}}

\usepackage{xspace}

\author{Cheng Chu}
\email{chu6@iu.edu}
\affiliation{%
  \institution{Indiana University Bloomington}
  \country{USA}}

\author{Nai-Hui Chia}
\email{naichia@iu.edu}
\affiliation{%
  \institution{Indiana University Bloomington}
  \country{USA}}

\author{Lei Jiang}
\email{jiang60@iu.edu}
\affiliation{%
  \institution{Indiana University Bloomington}
  \country{USA}}

\author{Fan Chen}
\email{fc7@iu.edu}
\affiliation{%
  \institution{Indiana University Bloomington}
  \country{USA}}

\begin{document}

\title{QMLP: An Error-Tolerant Nonlinear Quantum MLP Architecture using Parameterized Two-Qubit Gates}

\begin{abstract}
Despite potential quantum supremacy, state-of-the-art quantum neural networks (QNNs) suffer from low inference accuracy. 
First, the current Noisy Intermediate-Scale Quantum (NISQ) devices with high error rates of $10^{-3}$ to $10^{-2}$ significantly degrade the accuracy of a QNN. 
Second, although recently proposed Re-Uploading Units (RUUs) introduce some non-linearity into the QNN circuits, the theory behind it is not fully understood. 
Furthermore, previous RUUs that repeatedly upload original data can only provide marginal accuracy improvements.
Third, current QNN circuit ansatz uses fixed two-qubit gates to enforce maximum entanglement capability, making task-specific entanglement tuning impossible, resulting in poor overall performance. 
In this paper, we propose a Quantum Multilayer Perceptron (QMLP) architecture featured by
error-tolerant 
input embedding,
rich nonlinearity,
and enhanced variational circuit ansatz with parameterized two-qubit entangling gates.
Compared to prior arts, 
QMLP increases the inference accuracy on the 10-class MNIST dataset by 10\% with 2$\times$ fewer quantum gates and 3$\times$ reduced parameters.
Our source code is available and can be found in~\cite{QMLP_code}~\footnote{This paper has been accepted by the \textit{ACM/IEEE International Symposium on Low Power Electronics
and Design (ISLPED ’22)}.}.
\end{abstract}


\copyrightyear{2022} 
\acmYear{2022} 
\setcopyright{acmcopyright}\acmConference[ISLPED '22]{ACM/IEEE International Symposium on Low Power Electronics and Design}{August 1--3, 2022}{Boston, MA, USA}
\acmBooktitle{ACM/IEEE International Symposium on Low Power Electronics and Design (ISLPED '22), August 1--3, 2022, Boston, MA, USA}
\acmPrice{15.00}
\acmDOI{10.1145/3531437.3539719}
\acmISBN{978-1-4503-9354-6/22/08}


\maketitle

\section{Introduction}
\label{sec:introduction}
\textit{Quantum Computing} is a novel computing paradigm that solves classically intractable problems with substantially higher efficiency and speed. Due to the quantum parallelism and the effect of interference and entanglement, it has been demonstrated to have exponential advantage in various machine learning tasks~\cite{wang2021quantumnas, jiang2021co, tacchino2019artificial,tacchino2020quantum, havlivcek2019supervised, cong2019quantum}.
A quantum neural network (QNN)~\cite{biamonte2017quantum} is able to generate the correlation between variables that are inefficient to represent through classical computation by defining a feature map that maps classical data into the quantum Hilbert space.
Algorithms are developed for both inferring~\cite{schuld2014quest} and training~\cite{sweke2020stochastic} a quantum neural network.  
Although quantum advantage has been demonstrated, \textit{QNNs suffer from low inference accuracy}. State-of-the-art QNNs~\cite{jiang2021co,wang2021quantumnas} achieve $<$60\% accuracy when inferring a 10-class MNIST~\cite{mnist} dataset, i.e., the smallest and simplest benchmark in the classical machine learning domain. The low inference accuracy of state-of-the-art QNNs is caused by three factors, i.e., the Noisy Intermediate-Scale Quantum (NISQ) devices, the lack of nonlinearity, and the very limited learning capability of the QNN circuit ansatz.

First, quantum gates in the NISQ era are highly error-prone, and hence greatly decrease the inference accuracy of QNNs. Noises~\cite{Endo:JPSJ2021} introduced by imperfect fabrication, crosstalk, and non-ideal control and readout result in various types of errors, e.g., bit-flip and phase-flip errors~\cite{wang2021noise, funcke2020measurement}. 
Although recent work creates an automatic framework, QuantumNAS~\cite{wang2021quantumnas}, to search the most noise-resistant QNN within the design space of a large variational quantum circuit consisting of pre-defined parameterized quantum gates, QuantumNAS does not have any error awareness during model training. 
In another word, the framework just na\"ively searches the most accurate QNN within the design space that is not optimized for error tolerance, and then add noise during circuit mapping to retrain the selected model. 
Moreover, QuantumNAS uses the \textit{entanglement encoding} that may greatly degrade quality of the encoded quantum input feature maps (QIFPs) by propagating a single error to all parts of a QIFP. Such encoded input produces only low inference accuracy, since it is difficult for any type of networks to achieve high inference accuracy when low-level features are wrong.

Second, prior QNNs~\cite{wang2021quantumnas} do not have much nonlinearity due to the linear and unitary nature of quantum mechanics. A deep structure cannot help such models to improve their performance, since various nonlinear activations are a crucial factor for multi-layer deep neural networks to achieve high inference accuracy. Recent QNNs adopt data Re-Uploading Units (RUUs)~\cite{Perez:quatum2020} to gain nonlinearity by repeatedly applying their data encoding unitaries to a QNN. The orginal RUUs have demonstrated improved performance, but the theory behind is not fully understood. Moreover, prior RUUs just na\"ively re-upload identical QIFPs into a QNN, providing only limited amount of nonlinearity.

Third, the variational quantum circuit ansatz used in current QNNs utilize fixed two-qubit gates, i.e., \texttt{CNOT}, to force maximum entanglement between two qubits. The lack of trainablity on qubits entangling power limits the function space that a QNN can learn.
Therefore, it is difficult for QNNs to obtain the optimal mapping function, as in classical machine learning, to capture all potential correlations in the inputs.

In this paper, we propose a Quantum Multilayer Perceptron architecture, namely QMLP, to address the aforementioned challenges. Our main contributions can be summarized as follows:
\begin{itemize}[leftmargin=*, topsep=0pt, partopsep=0pt]
\item \textbf{An error-tolerant network architecture}. 
We propose a new network architecture featuring an error-tolerant encoding method by which a QMLP encodes each input qubit separately, confining errors on the encoding circuit to only a portion of the encoded QIFPs. 
In this way, more accurate low-level features can be generated for processing by later QMLP components.
\item \textbf{A re-uploading layer with adjustable nonlinearity.}
We investigate the theory behind the data Re-Uploading Units~\cite{Perez:quatum2020} and further present advanced RUU layer designs that are capable to provide tunable nonlinearity. Compared to original RUUs, our new RUU layers can simulate various types of nonlinearity that pave the way for constructing \textit{deep} QNNs.
\item \textbf{A circuit ansatz with enhanced trainability.} 
We present a circuit ansatz using parameterized two-qubit gates, which applies appropriate entanglement between two qubits through model training, making it easier to learn optimal QNN parameters.
\item  \textbf{High inference accuracy}. We implemented, evaluated, and compared QMLP against prior QNNs such as QuantumFlow~\cite{jiang2021co} and QuantumNAS~\cite{wang2021quantumnas}. Compared to prior QNN designs, QMLP increases the inference accuracy on the 10-class MNIST dataset by 10\% with 2$\times$ fewer quantum gates and 3$\times$ reduced parameters.
\end{itemize}

\section{Background and Motivation}
\label{sec:background}

\subsection{Quantum Basics}
\textbf{Qubits}. 
Unlike a conventional bit, a quantum bit (qubit)~\cite{sweke2020stochastic} can be in a linear combination of the two basis states 0 and 1: $|\phi\rangle=\alpha |0\rangle + \beta |1\rangle$, for $\alpha,\beta\in \mathbb{C}$, and $|\alpha|^2+|\beta|^2=1$. A ``superposition'' of basis states enables an $n$-qubit system to represent a linear combination of $2^n$ basis states. On the contrary, a classical $n$-bit register file can store only one of these $2^n$ states.

\begin{figure}[t!]
    \centering{\includegraphics[width=1\linewidth]{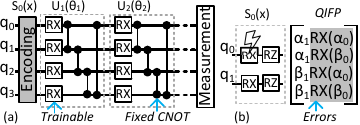}}
    \vspace{-18pt}    
    \caption{A standard QNN. (a) The circuit ansatz; (b) The error propagation in a noisy entanglement encoding circuit.}
    \label{f:vqc_background}\vspace{-10pt}
\end{figure}

\textbf{Quantum Neural Networks}. 
A universal QNN architecture based on trainable \textit{variational quantum circuits} (VQCs)~\cite{biamonte2017quantum} is shown in Figure~\ref{f:vqc_background} (a).
Given a classical training set $\mathcal{T}=(\mathbf{x_i}, y_i)$ of $n$-dimensional feature map vectors $\mathbf{x_i}=(x_i^0, x_i^1,{\ldots}x_i^{n-1})$ and its label $y_i$, a \textit{quantum encoding layer}, $S_0(\mathbf{x_i})$, is applied to the ground quantum state $|{\phi}_0\rangle=|0\rangle^{{\otimes}N}$, to generate the \textit{quantum input feature map} (QIFP) $|{\phi}_1\rangle$.
A parameterized VQC ansatz is then utilized to manipulated the QIFP through unitary transformation, i.e., $|\phi_1\rangle\rightarrow U(\theta)|\phi_1\rangle$, where $\theta$ is a set of free variables for adaptive optimizations. 
The VQC ansatz consisting of a sequence of \textit{quantum gates} fundamentally defines the types of functions that a QNN can compute.
Similar to the ``layers'' that form the architecture of a classical neural network, a QNN can be constructed by stacking multiple layers of a VQC ansatz. i.e., $U(\theta)=U_L(\theta_L){\cdots}U_2(\theta_2)U_1(\theta_1)$. 
The predicted classification can be obtained by a quantum state \textit{measurement}, followed by necessary classical \textit{post-processing}. A loss function is predefined to calculate the difference between the output of the QNN and the true target value $y_i$. The training of a QNN iteratively search for the best parameters in a VQC via a hybrid quantum-classical optimization procedure.

\textbf{QNN Circuit Ansatz}.
VQC ansatzes are empirically designed based on prior knowledge of the target problem, as it is often impossible to clearly predict and explain how different circuit ansatzes will perform on various datasets. 
The circuit ansatz shown in Figure~\ref{f:vqc_background} (a), consisting of parameterized single-qubit rotation gates (i.e., \texttt{RX}, \texttt{RY}, \texttt{RZ}, or a combination thereof) followed by nearest-neighbor coupling of all qubits using fixed two-qubit \texttt{CNOT} gates, has demonstrated superior prediction accuracy compared to other counterparts and is adopted as a basic VQC ansatz in various classification tasks~\cite{harrow2009quantum, biamonte2017quantum, wang2021quantumnas}. 
The central idea of such designs is that single-qubit rotations provide a way to parameterize circuits, while two-qubit gates entangle the circuit.

\textbf{Errors on NIST Machines}. 
In quantum computing, noises happen due to the imperfect control signals, unavoidable crosstalk interactions between qubits, or interference from the environment~\cite{Endo:JPSJ2021}. NISQ devices are severely limited in circuit width and depth because they suffer from \textit{decoherence error} over time, making it impractical to implement complex quantum circuits. In addition, quantum gates introduce \textit{operation errors}, e.g., bit-flip errors and phase-flip errors~\cite{wang2021noise, funcke2020measurement}, into the quantum machine. 
As a result, a quantum machine needs to be characterized and calibrated frequently to mitigate noise impacts.

\begin{figure*}[t]\centering
\begin{minipage}{0.49\textwidth}
    \center{\includegraphics[width=\linewidth]{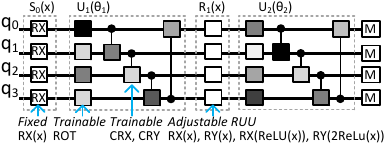}}\vspace{-10pt}
    \caption{The QMLP architecture.}\vspace{-8pt}
    \label{f:QMLP_overview}
\end{minipage}\hspace{2pt}
\begin {minipage}{0.24\textwidth}
    \center{\includegraphics[width=\linewidth]{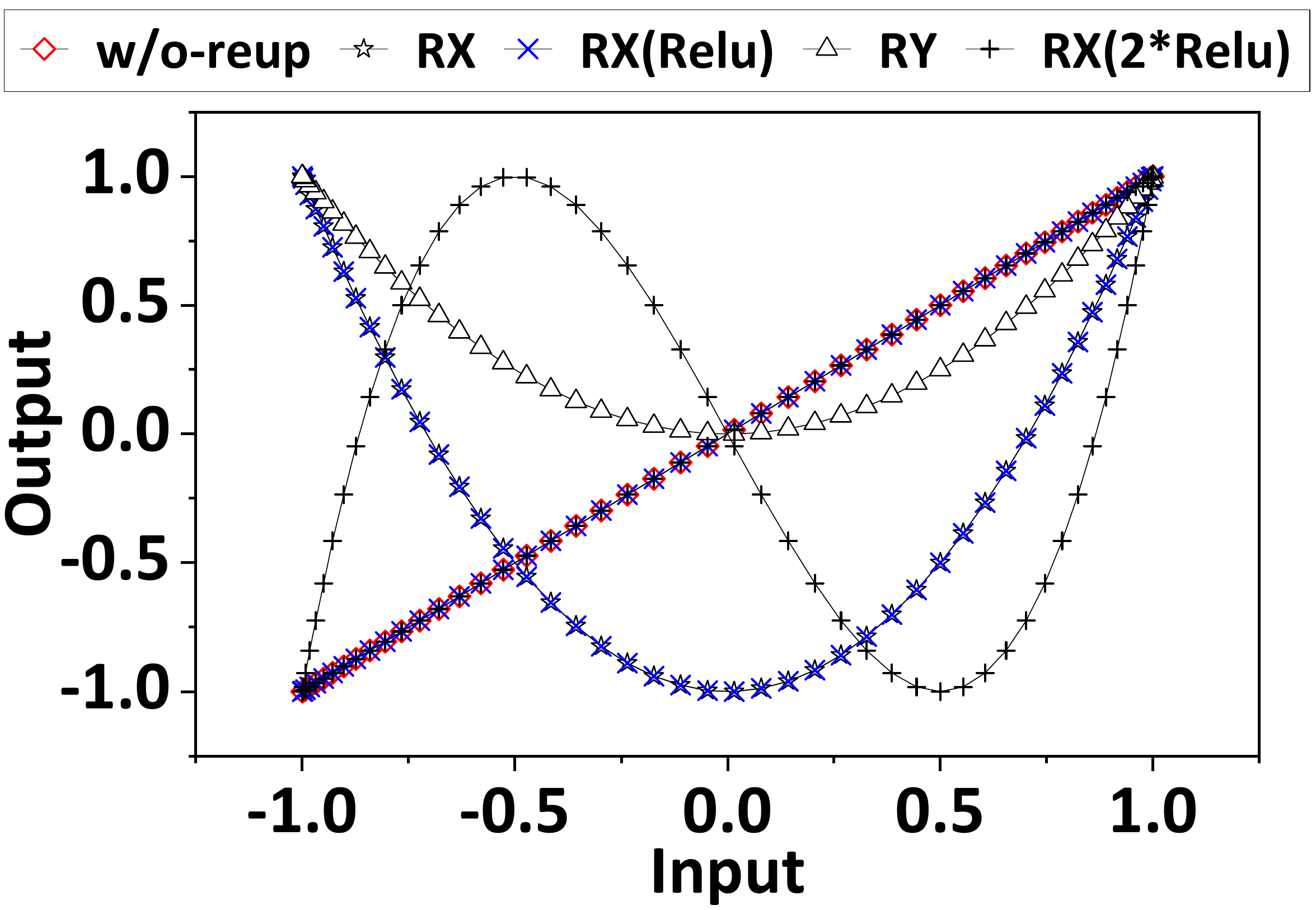}}\vspace{-12pt}
    \caption{Adjustable RUUs with rich nonlinearities.}\vspace{-8pt}
    \label{f:ruu_nonlinearity}
\end{minipage}\hspace{6pt}
\begin {minipage}{0.24\textwidth}
    \center{\includegraphics[width=\linewidth]{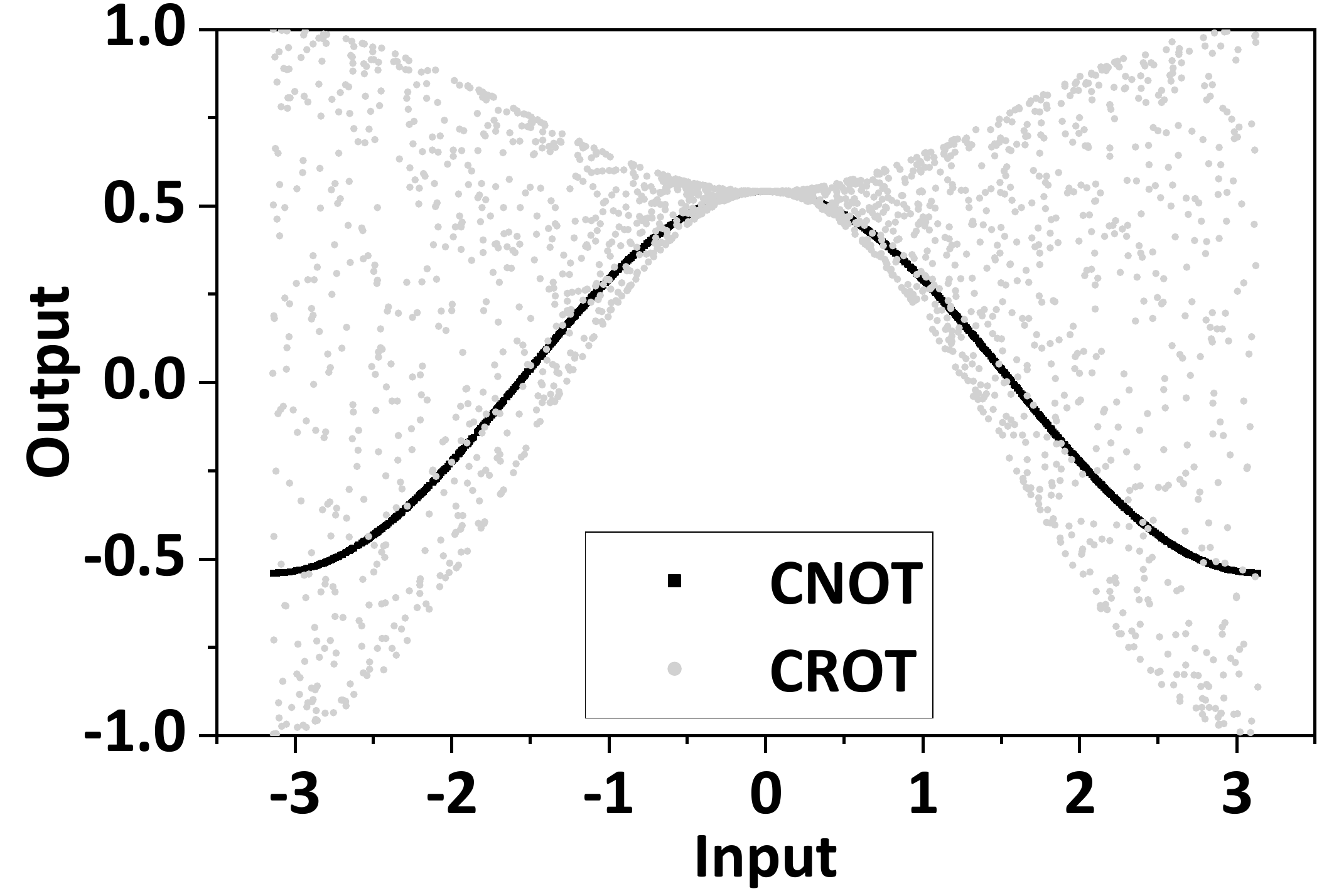}}\vspace{-12pt}
    \caption{Comparison between VQCs with \texttt{CNOT} and \texttt{CROT} gates.}\vspace{-8pt}
    \label{f:ruu_cnot_crot_cmp}
\end{minipage}
\vspace{-6pt}
\end{figure*}


\subsection{Quantum Data Encoding}
Recent QNNs such as QuantumNAS~\cite{wang2021quantumnas} prepares the quantum input feature maps (QIFPs) by encoding classical data into the amplitudes of a quantum state, e.g., $|{\phi}_1\rangle=x_i^0|00\rangle+x_i^1|01\rangle+x_i^2|10\rangle+x_i^3|11\rangle$ for a 4-dimensional input vector as shown in Figure~\ref{f:vqc_background} (b). Since the information contained in the classical input is entangled in the quantum system and embedded in each encoded qubit, we refer it as \textit{entanglement encoding} in this paper. The efficiency of such entanglement encoding is higher than other encoding schemes as it only requires $log_2(n)$ qubits to represent an $n$-dimensional input.

\textbf{QIFP Quality}.
Input data encoding is crucial to the overall design and performance of a QNN because it is impossible to learn an accurate model and hence achieve high inference accuracy using inaccurate quantum input feature maps.
While in the current NISQ era, it is very likely that an error may happen on the quantum gates used in the encoding layer. The entanglement encoding, however, will propagates \textit{errors on even a single gate} to \textit{all parts} of a QIFP as concepturally illustrated in Figure~\ref{f:vqc_background} (b). 
Furthermore, despite the benefits for encoding efficiency, such entanglement encoding is achieved with complex quantum state preparation circuitry consisting a sequence of quantum rotation gates, e.g. two \texttt{RY} gates, one \texttt{RX} gate, and one \texttt{RZ} gate per qubit in~\cite{wang2021quantumnas}, resulting $>$30\% hardware overhead and increased error rates. 
Therefore, it is important to investigate a new quantum encoding architecture that is easy to implement and capable of mitigating the negative impact of noisy encoding circuits on QIFP quality.



\textbf{Input Re-Uploading.}
Following the analogy of classical neural networks, \textit{deep} QNNs can be constructed by casting a parameterized VQC ansatz multiple times in a structure with hidden layers.
However, the obstacle that arises is how to achieve nonlinearity in the hidden units with unitary quantum circuits. 
Prior works~\cite{Perez:quatum2020, havlivcek2019supervised} proposed an input Re-Uploading Unit, where the quantum encoding layer, $S_0(\mathbf{x_i})$, are repeatedly applied before each parameterized VQC ansatz.
This method outperforms the default variational quantum algorithm on model expressiveness and final prediction accuracy because, as the authors observed -- ``it introduces nonlinear behavior", but the theory behind input re-uploading is not fully understood. 
{Therefore, it is desirable to provide a technical discussion and explain why and how an RUU provides nonlinearity. Furthermore, a generic RUU with flexible and tunable nonlinearity will potentially be beneficial to implement very wide and deep QNN circuits.}

\subsection{Related Work}
Multiple QNNs~\cite{jiang2021co, wang2021quantumnas, tacchino2019artificial,tacchino2020quantum, havlivcek2019supervised, cong2019quantum} are proposed to demonstrate the quantum supremacy by predicting the correlation between variables that are inefficient to represent through classical computation. 
QuantumFlow~\cite{jiang2021co} realizes the functionality of a neuron by non-trainable quantum gates. By using binary weights, the non-trainable-gate-based network is supposed to be more tolerant of errors. However, its high circuit complexity (e.g., 1114 quantum gates) negates the potential robustness advantage and conversely leads to $\sim$10\% inference accuracy on 10-class MNIST classification when noise is considered.
Most QNNs~\cite{wang2021quantumnas, tacchino2019artificial,tacchino2020quantum, havlivcek2019supervised, cong2019quantum} use and train variational quantum circuits to achieve their neural network function.
Specifically, QuantumNAS~\cite{wang2021quantumnas} creates a search framework to automatically build a sub-variational-quantum-circuit that can achieve state-of-the-arts 10-class MNIST classification accuracy on NISQ devices using different circuit components inherited from a big SuperCircuit. However, neither the large SuperCircuit or any of its sub-circuit has error tolerance awareness during model training (i.e., circuit noise is only considered during subsequent mapping and retraining process) or implements strong nonlinearity, limiting further performance improvement and system scalibility. 

\section{The QMLP Architecture}
\label{sec:design}
Figure~\ref{f:QMLP_overview} shows the overview of the proposed QMLP architecture and its relevant components. In this example we consider a two-layer QMLP with four qubits. The number of layers and qubits in a generic QMLP circuit can be adjusted to fit the problem of interest.
QMLP prepare the quantum input feature maps using a linear encoding layer, $S_0(\mathbf{x})$, consisting only \texttt{RX} gates.
Similar to the architecture of a classical neural network, QMLP is implemented from multiple layers by stacking a parameterized VQC ansatz several times.
The circuit ansatz, $U_i(\mathbf{\theta_i})$, proposed in this work is implemented as parameterized single-qubit rotation gates (i.e., \texttt{ROT}) on each qubit followed by nearest-neighbor coupling of all qubits using parameterized two-qubit gates (i.e., \texttt{CRX, CRY}).
The parameter matrix $\mathbf{\theta}_i$ are subject to iterative optimization, which can be seen as the weights in classical neural networks. 
For all layers except the input one, we apply an re-uploading unit, $R_i(\mathbf{x})$, which provides adjustable nonlinearity to the QMLP. 
We measure each qubits on Pauli-Z basis and obtain a classical value in the range of [-1, 1]. We then obtain the 10-class probability by using a small fully-connected layer for post-processing.






\subsection{Input Data Encoding and Re-Uploading}
QMLP encodes the inputs into the rotation angles of qubits by using single Pauli-rotation gates.
The encoding scheme can be mathematically formulated as,
{

\begin{equation}
\label{eq:encoding}
S(\mathbf{x_i}) = e^{-i{\frac{x^0_i}{2}}{\sigma_j}}{\otimes\dots\otimes}e^{-i{\frac{x^{n-1}_i}{2}}{\sigma_j}}
\end{equation}
\vspace{-16pt}
}

\noindent
where $\sigma_j\in\{\sigma_x, \sigma_y, \sigma_z\}$, $\mathbf{x_i}=(x_i^0, x_i^1,{\ldots}x_i^{n-1})$ represents the \textit{n} dimensional input data, 
and $\otimes$ represents the tensor product operation which combines the state spaces of all the individual qubits.
Compared with entanglement encoding scheme~\cite{wang2021quantumnas}, 
our scheme confines each single input information to a distinct qubit.
This forced isolation minimizes the impact of errors on encoding circuits, 
thereby maximizing the fidelity transformation between classical input data and quantum input feature maps.
In particular, the linear representation in Equation~\ref{eq:encoding} does not necessarily imply that the gate or function it computes is linear. 
Below we show how the generic encoding scheme is adjusted for 
\textit{linear input embedding} and \textit{non-linear input re-uploading}.

\begin{figure}[t!]
    \centering \vspace{0pt}
    \includegraphics[width=1\linewidth]{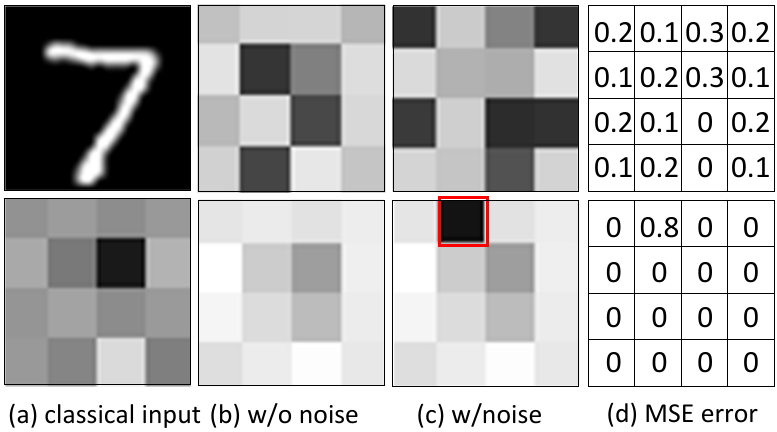}
    \vspace{-20pt}    
    \caption{A comparison between entanglement encoding and our method under a single gate bit-flip error: (a) Original classical MNIST image (top) and downsampled image (bottom); (b-c) The QIFPs generated with ideal and noisy encoding circuits using entanglment encoding (top) and our method (bottom); (d) The pixel-wise mean-square error (MSE) between column (a) and (c).}
    \label{f:encoding_noise_impact}
    \vspace{-18pt}
\end{figure}

\textbf{Error-Tolerant Linear Input Embedding.}
QMLP applies single-qubit rotation gate \texttt{RX} on the ground state $|{\phi}_0\rangle=|0\rangle^{{\otimes}N}$, where the rotation angle is determined by the classical inputs. There are two main reasons for choosing this encoding scheme.
First, such \textit{linear feature-embedding circuit} ensures that the relative value/distance of the input data remains unchanged, thereby preserving all the correlation in the original data to the greatest extent.
Second, we reduce the complexity of the encoding circuit to one \texttt{RX} gate per qubit. Compared to the $\sim$30\% encoding circuit overhead in QuantumNAS~\cite{wang2021quantumnas}, our encoding scheme greatly avoids the adverse effects of noise in complex circuits, resulting improved QIFP quality.
The comparison between \textit{entanglement encoding} used in QuantumNAS~\cite{wang2021quantumnas} and our encoding methods is visually illustrated in Figure~\ref{f:encoding_noise_impact}. With ideal quantum encoding circuits, the quantum input feature maps generated by both methods preserve the correlation exhibited in the original data as shown in Figure~\ref{f:encoding_noise_impact} (a) and (b). We then assume a bit-flip error occurs on only one of the \texttt{RX} gates. As shown in the top row of Figure~\ref{f:encoding_noise_impact} (c) and (d), the \textit{single} error is propagated in the \textit{whole QIFP} when entanglement encoding is applied. In contrast, our method ensures that the error is confined within a limited spatial region in the QIFP, thereby resulting strong error tolerance. 
It is worth mentioning that our encoding method also improves the overall noise immunity of QNNs as it reduces potential noise propagation paths due to information decoupling between qubits.
The system effectiveness is demonstrated in model accuracy results reported in Section~\ref{sec:results}.


\textbf{Nonlinear Input Re-Uploading}.
The empirical RUU configuration in~\cite{Perez:quatum2020} is to apply identical input encoding, e.g., $S_0(\mathbf{x})$ in Figure~\ref{f:QMLP_overview}, before each trainable layer, i.e., $U_i(\theta_i)$, to introduce nonlinearity into a unitary quantum circuit. 
Below we provide a technical discussion and explain why and how input re-uploading provides nonlinearity. We then demonstrate a generic re-uploading unit with flexible and tunable nonlinearity. 

Equation~\ref{eq:encoding_reuploading} denotes the quantum state evolution with the original identical RUUs applied in each layer. Different from the linear quantum state prepared by a VQC without RUUs, i.e., $|\psi(x,\theta)\rangle =U(\theta)|\phi_1\rangle$, it is clear that input re-uploading produces higher-order terms of the quantum input feature map, thereby enabling the quantum system to 
perform nonlinear functions with respect to the input.
For ease of understanding, we show a special case of Equation~\ref{eq:encoding_reuploading} in Equation~\ref{eq:encoding_simple_reuploading} by forcing the VQC circuit function in each layer to be identity transformation. 

{\small
\vspace{-0.05in}
\vspace{-0.05in}
\begin{equation}\label{eq:encoding_reuploading}
\begin{split}
|\psi(x,\theta)\rangle 
& = {U_L}(\theta_L){S_0}(\mathbf{x})\cdots{U_2}(\theta_2){S_0}(\mathbf{x}){U_1}(\theta_1){S_0}(\mathbf{x})|\phi_0\rangle \\
& ={U_L}(\theta_L){S_0}(\mathbf{x})\cdots{U_2}(\theta_2){S_0}(\mathbf{x}){U_1}(\theta_1)|\phi_1\rangle 
\end{split}
\end{equation}
\vspace{-0.1in}
\begin{equation}\label{eq:encoding_simple_reuploading}
\begin{split}
|\psi(x,\theta)\rangle 
& = {I_L}(\theta_L){S_0}(\mathbf{x})\cdots{I_2}(\theta_2){S_0}(\mathbf{x}){I_1}(\theta_1){S_0}(\mathbf{x})|\phi_0\rangle \\
& ={(S_0(\mathbf{x}))}^{L-1}|\phi_1\rangle 
\end{split}
\end{equation}
\vspace{-0.1in}
}

Based on the theoretical study and motivated by the original RUU designs~\cite{Perez:quatum2020,havlivcek2019supervised}, we set out to explore and design RUUs, denoted as $R_i(\mathbf{x})$ in Figure~\ref{f:QMLP_overview}, that can provide more flexible and tunable nonlinearities. Instead of applying identical input re-uploading, we evaluated RUUs with single-qubit rotation gates, i.e., \texttt{RX}($\mathbf{x}$), \texttt{RY}($\mathbf{x}$), \texttt{RZ}($\mathbf{x}$), classical nonlinear pre-processing, e.g., \texttt{ReLU}($\mathbf{x}$), 
or combination of them, e.g., \texttt{RX(ReLu}($\mathbf{x}$)). 
Figure~\ref{f:ruu_nonlinearity} shows several examples of nonlinear functions when different ${R_i}(\mathbf{x})$ are applied in a two-layer network with only one qubit.
In practical multilayer QNNs, the nonlinearity introduced by RUU is related to the input encoding method and circuit structure. Therefore, specific adjustment and evaluation of the RUU structure should be carried out.

\subsection{Enhanced VQC with Parameterized Two-Qubit Gates}
Previous QNN circuit ansatz~\cite{harrow2009quantum, biamonte2017quantum, wang2021quantumnas} is implemented with parameterized single-qubit gates to provide circuit trainability and fixed two-qubit \texttt{CNOT} gates to entangle the circuit with maximum entanglement power.
In this work, we explore trainable entanglement by replacing fixed \texttt{CNOT} gates with parameterized two-qubit gates, hoping that adaptive and flexible entanglement capabilities can be better utilized for QNN algorithms on different tasks.
For practical candidate gates, we consider two commonly used parameterized two-qubit gates, i.e., \texttt{CRX($\theta$)} and \texttt{CROT($\phi, \theta, \omega$)}. Our preliminary results shows that \texttt{CRX($\theta$)} gates outperforms \texttt{CROT($\phi, \theta, \omega$)} gates in terms of final classification accuracy, hardware overhead, and training complexity. 
Below we briefly explain the case when \texttt{CNOT} is replaced by its parameterized version, \texttt{CRX($\theta$)}. 
\texttt{CNOT} and \texttt{CRX($\theta$)} are defined by the following matrices.
The entangling power of \texttt{CRX($\theta$)} increases 
between the value of $\theta$ from 0 to $\pi$.  
For ${\theta}=0$, a \texttt{CRX} gate is an identity gate, 
while for ${\theta}=\pi$ it implement a fixed \texttt{CNOT} gate. 

\textit{Note: When preparing for the final submission of this paper, we
were made aware that the effects of parameterized two-qubit gates in the variational quantum eigensolvers has already been studied in~\cite{rasmussen2022parameterized}.
While there is conceptual overlap between this work and ours, 
we are studying different candidate gates,
different circuit ansatz, 
and targeting a different quantum algorithm.
}

{\footnotesize
\[
\texttt{CNOT} =
\begin{bmatrix}
    1  &  0 &  0 &  0     \\
    0  &  1 &  0 &  0     \\
    0  &  0 &  0 &  1     \\
    0  &  0 &  1 &  0     
\end{bmatrix}\hspace{0.05in}
\texttt{CRX}(\theta)= 
\begin{bmatrix}
    1  &  0 &  0 &  0     \\
    0  &  1 &  0 &  0     \\
    0  &  0 &  cos(\frac{\theta}{2}) &  -i{\cdot}sin(\frac{\theta}{2})     \\
    0  &  0 &  -i{\cdot}sin(\frac{\theta}{2}) &  cos(\frac{\theta}{2})     
\end{bmatrix} 
\]
}

To evaluate the performance of the VQC ansatz with different parameterized two-qubit gates, i.e., $U_i(\theta_i)$ in Figure~\ref{f:QMLP_overview}, we implemented a simple two-layer QMLP with two qubits. We sweep the classical value that is encoded in one qubit and report how the learned function space changes subject of the sweep in Figure~\ref{f:ruu_cnot_crot_cmp}. By relaxing the forced maximum entanglement, it shows that VQCs with parameterized two-qubit gates are able to learn mapping functions in an enlarged search space, indicating that QNNs constructed from such circuit ansatz is also likely to learn a more accurate model.


\begin{table*}[bhpt]\centering
\caption{Different QMLP schemes.}\label{tab:qmlp_config}\vspace{-12pt}
\begin{tabular}{|c||c|c|c|c|c|c|c|}\hline
Schemes          & Re-upload Gate             & Qubit \#     & 2-qubit Gate   & Parameterized Gate.    & Total Gates \#    & Total Param. \#   & Block  \#         \\\hline  \hline
RX-CNOT         &RX                         &16         &CNOT              &ROT                     &94                 &96                 &2\\\hline
DEEP-RX-CNOT    &RX                         &16         &CNOT              &ROT                     &188                &192                &4\\\hline
RX-CRX          &RX                         &16         &CRX               &ROT, CRX                &96                 &128                &2\\\hline
DEEP-RX-CRX     &RX                         &16         &CRX               &ROT, CRX                &192                &256                &4  \\\hline
RXY(Relu)-CRX  &RX, RY(2$\times$RELU(x))   &16         &CRX               &ROT, CRX                &96                 &128                &2 \\\hline
RXX(Relu)-CRX  &RX, RX(RELU(x))            &16         &CRX               &ROT, CRX                &96                 &128                &2  \\\hline
RXY-CRXY       &RX, RY                     &16         &CRX, CRY          &ROT, CRX, CRY           &96                 &128                &2   \\\hline
RXY-CNOT       &RX, RY                     &16         &CNOT              &ROT                     &94                 &96                 &2  \\\hline
\end{tabular}\vspace{-8pt}
\end{table*}

\begin{table*}[bpht] \centering
\caption{Design overhead comparison.}\label{tab:overhead}\vspace{-12pt}
\begin{threeparttable}
\begin{tabular}{|c||c|c|c|c|c|c|c|}
\hline
Design          & 1-qubit Gate \#  & 2-qubit Gate \# & Param. \#    & Depth     & Qubit \#     & Input Size     & Gate List         \\\hline  \hline
QuantumFlow1~\cite{jiang2021co}  &\multicolumn{2}{c|}{386$^*$}            & N/A      &           &256        &16 $\times$ 16 &RY, H, X, CONT, CZ   \\\hline
QuantumFlow2~\cite{jiang2021co}  &\multicolumn{2}{c|}{1114$^*$}           & N/A      &           &64         &8 $\times$ 8   &RY, H, X, CONT, CZ   \\\hline

QuantumNAS~\cite{wang2021quantumnas}     &116(80)            &80              &480       &88         &10         &6 $\times$ 6   &RX, RZ, RY, CU3, U3  \\\hline
QMLP            &64(32)             &32              &128       &36         &16         &4 $\times$ 4   &RX, ROT, CRX          \\\hline
\end{tabular}
  
\begin{tablenotes}[para,flushleft]
\textit{$^*${\cite{jiang2021co} only reported the total number of gates.}}
\end{tablenotes}
\end{threeparttable}\vspace{-16pt}
\end{table*}

\subsection{Scalability of the QMLP Architecture }
\begin{figure}[h]\vspace{-12pt}
\setlength{\abovecaptionskip}{-4pt}
\setlength{\belowcaptionskip}{0pt}
    \center{\includegraphics[width=1\linewidth]{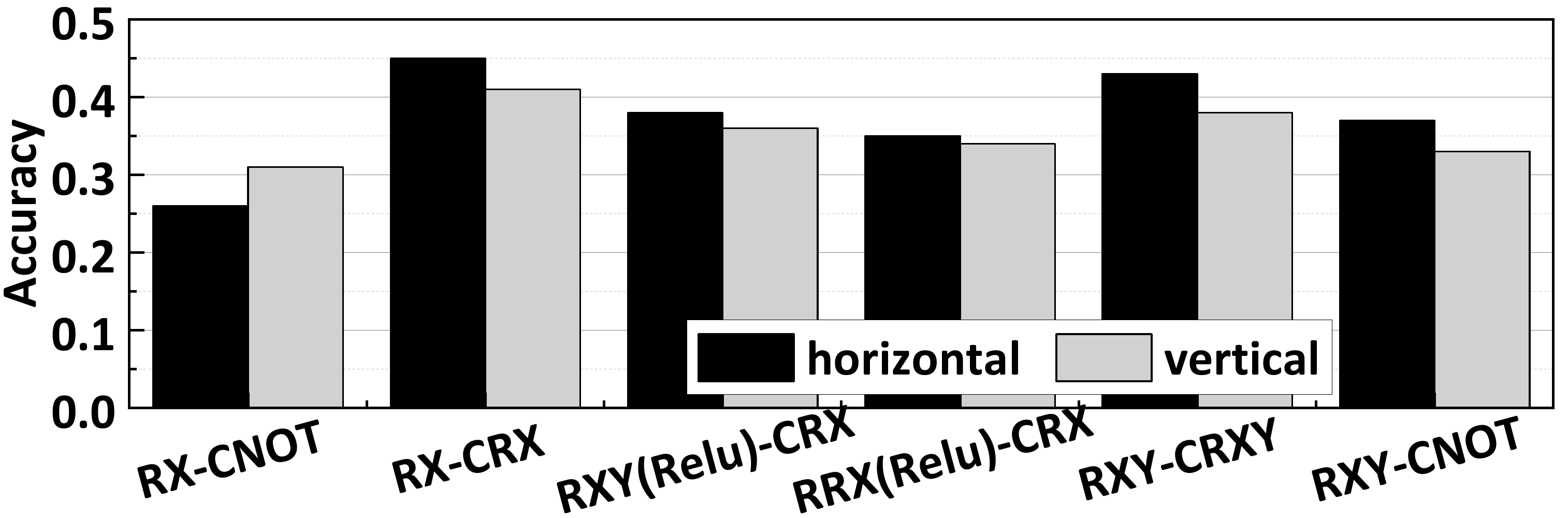}}\vspace{6pt}
    \caption{The comparison between vertical QMLP and horizontal QMLP.}
    \label{f:vertical_horizontal}\vspace{-10pt}
\end{figure}

Without RUU, nonlinearity can only be introduced through input encoding (and output measurement that is not studied in this work), and subsequent evolution is linear. 
RUU guarantees the nonlinearity of the quantum feature map after each processing layer in the Hilbert space of the quantum system.
We demonstrate deep QMLP models can be effectively implemented with our proposed adjustable RUUs due to the increased expressivity introduced by the nonlinear data reuploading.  
In addition, recent work~\cite{schuld2021effect} has shown that implementing deep QNNs with multiple RUUs vertically or horizontally produces no difference on model accuracy.
This observation provides a practical implementation for deep QNNs by trading depth for width. We implemented the two-layer QMLP in Figure~\ref{f:QMLP_overview} in both vertical and horizontal ways and train both models with same configuration. We compare their inference accuracy in Figure~\ref{f:vertical_horizontal}. The two models demonstrated similar performance with $<$5\% qualitative accuracy difference. The slight distinction in performance is mainly caused by the difference in trainable circuit and measurement.  In practice, the implementation of a deep QMLP can be adjusted according to the available hardware resource, i.e., number of qubits and circuit depth, and requirement on error-tolerance.

\section{Experimental Methodology}
\label{sec:methodology}
\textbf{Benchmarks}.
We use MNIST~\cite{mnist} dataset on 10-class classification to evaluate QNN performance. Since quantum computing resources are limited and simulation time increase exponentially with the number of qubits, we down-sampled the 28$\times$28 MNIST images to  4$\times$4, similar to prior work~\cite{wang2021quantumnas, jiang2021co}.
The whole MNIST dataset are used for noise-free training and testing on a simulator.
We select the first 1000 images as testing data on real quantum processors.

\textbf{Simulation}. 
We build QMLP circuits using PennyLane~\cite{pennylane}, a quantum computing software library. The PennyLane APIs transform quantum circuits into torch layers, which can be used as \texttt{torch.nn.Sequential} modules and then simulated in PyTorch~\cite{pytorch}.
The trainable parameters in QMLP are converted as the weights of torch layers.
We use the \texttt{ADAM} optimizer with a weight decay of 0.0001. 
We set the training rate, batch size, and epoch respectively as 0.001, 32, and 30.
The source code for this paper is open and can be found in~\cite{QMLP_code}.

For performance comparison, 
we evaluate QMLP against the QuantumNAS~\cite{wang2021quantumnas} and QuantumFlow~\cite{jiang2021co}. We retain the original configuration (e.g., classical pre- and post-processing, circuit organization, etc.) of QuantumNAS and QuantumFlow, but adapt them to the same task (i.e., 10-class instead of the 4-class MNIST classification in~\cite{jiang2021co}). 
For QMLP design space exploration, we implement 8 QMLP circuits with various RUU configuration and different types of two-qubit gates in the trainable circuit ansatz. Details of the 8 QMLP schemes are summarized in Table~\ref{tab:qmlp_config}.

\textbf{Noise modeling}. 
Quantum errors on NISQ devices that most significantly impact circuit performance are bit-flip and phase-flip~\cite{wang2021noise, funcke2020measurement} that respectively erroneously recording an outcome as $\beta|0\rangle+\alpha|1\rangle$ or $\alpha|0\rangle-\beta|1\rangle$ given it was actually $\alpha|0\rangle+\beta|1\rangle$, and vice versa. The bit-flip and phase-flip errors can be equivalently modeled as probabilistic insertion of \texttt{X} or \texttt{Z} gates in a quantum circuit~\cite{funcke2020measurement}. In our model, we set the probability of phase-flip to 1\%. We set single-bit bit-flip to 1\%, and the bit-flip probability on a two-qubit gate is obtained based on the four probabilistic output combination, i.e., $99\% \times 99\%$, $99\% \times 1\%$, $1\% \times 99\%$, $1\% \times 1\%$ respectively.
We validate our noise model with \texttt{ibmq\_quito}~\cite{ibmq}, a 5-qubit IBM quantum processor, and confirm that the model is reasonably accurate in approximating gate errors.

\section{Results and Analysis}
\label{sec:results}
\begin{figure}[t]
    \center{\includegraphics[width=1\linewidth]{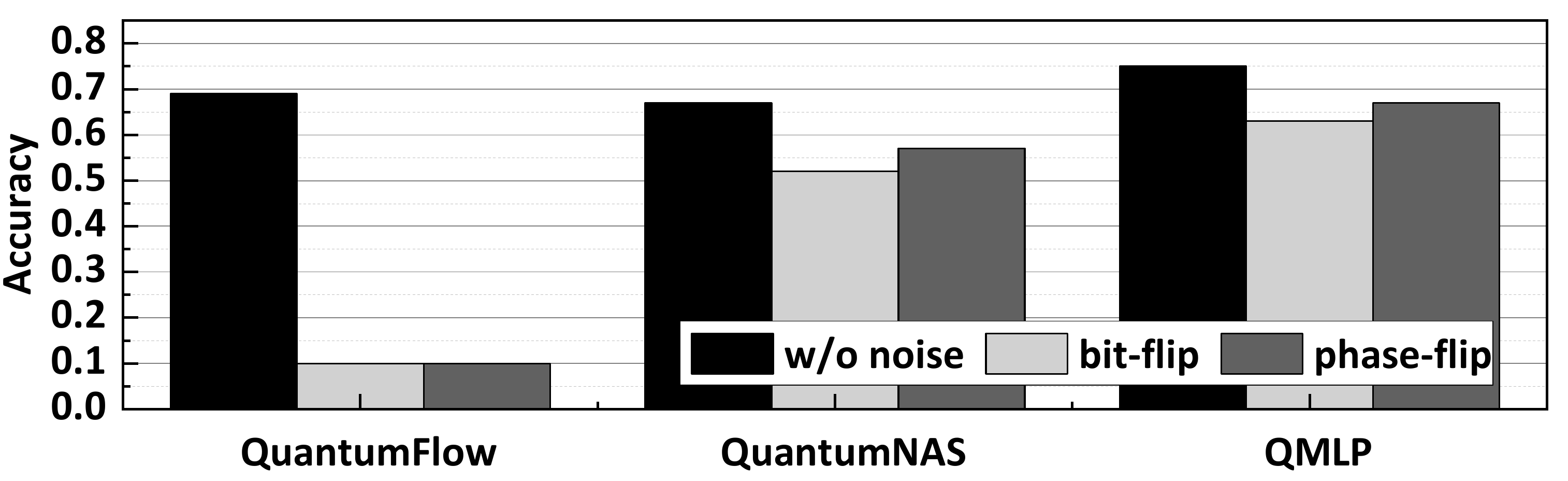}}\vspace{-12pt}
    \caption{The comparison between different QNNs.}
    \label{f:rank}\vspace{-20pt}
\end{figure}

\textbf{Comparison on QNN inference accuracy}.
We compare the inference accuracy of QMLP with the best performance of QuantmFlow~\cite{jiang2021co} and QuantumNAS~\cite{wang2021noise}. 
For QuantumFlow, we adopt their best design with a large 16$\times$16 MNIST image size.
For QuantumNAS, we report its highest accuracy by using the SuperCircuit.  
The QMLP model reported here is the default QMLP model, i.e., \texttt{RX-CRX} in Table~\ref{tab:qmlp_config}. 
As shown in Figure~\ref{f:rank}, QuantumFlow, QuantumNAS, and QMLP respectively achieved 69\%, 67\%, and 75\% accuracy on 10-class MNIST classification in the noise-free simulation.
When gate errors are considered, the accuracy on QuantumFlow drops sharply to 10\%, equivalent to random guessing, mainly due to its noise-unawareness and highly complex circuit with $>1000$ quantum gates. 
QuantumNAS delivers 52\% and 57\% inference accuracy with bit-flip and phase-flip errors are considered. In contrast, QMLP can mitigate the negative effect of gate errors, demonstrating a respective 63\% and 67\% inference accuracy.

\textbf{Comparison on QNN design overhead}.
Table~\ref{tab:overhead} compares the design overhead of QuantumFlow~\cite{jiang2021co}, QuantumNAS~\cite{wang2021noise}, and QMLP. Note that QuantumFlow provides difference  implementations based on different input sizes. We report hardware cost for two representative designs. 
Compared with QuantumFlow, QuantumNAS and QMLP significantly reduce the circuit complexity. 
QMLP uses more qubits than QuantumNAS as the \textit{entanglement encoding} in QuantumNAS uses a more compact data representations than our method. However, such an encoding circuit requires more quantum gates. Furthermore, the basic circuit ansatz used in QuantumNAS is more complex in both gate count and trainable parameters. Overall, QMLP reduces the number of gates and parameters by respectively 2$\times$ and 3$\times$ compared to QuantumNAS.


\begin{figure}[t]
    \center{\includegraphics[width=1\linewidth]{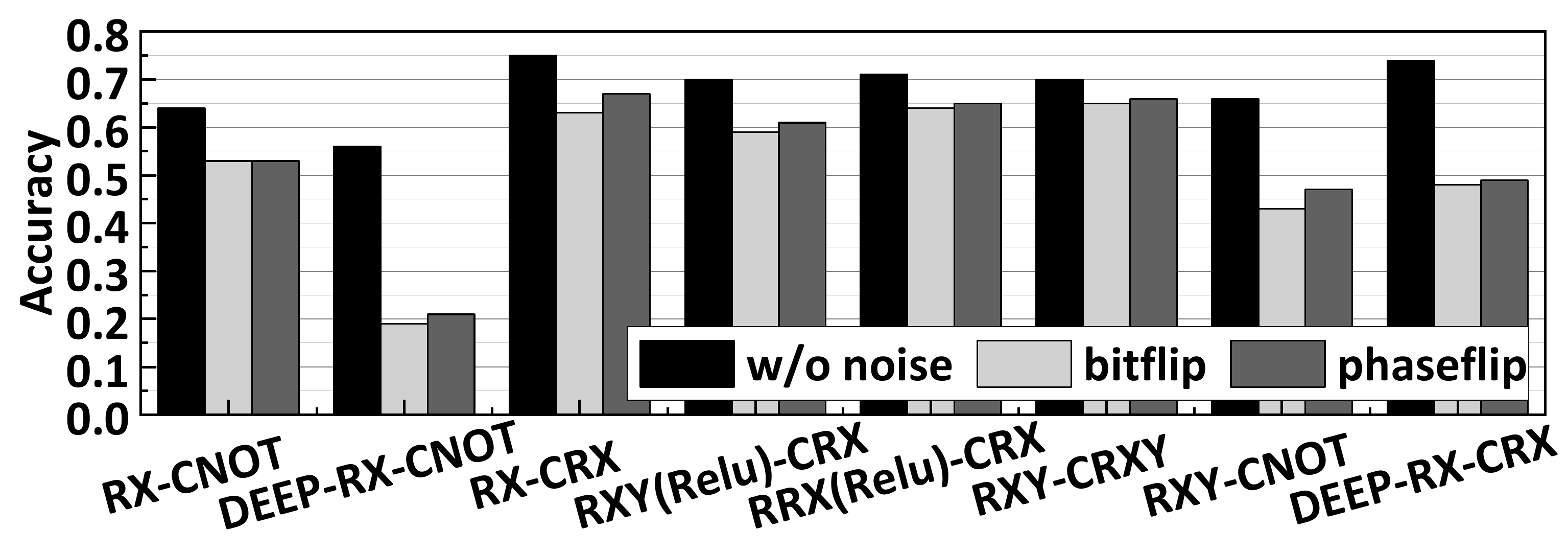}}\vspace{-12pt}
    \caption{The comparison between different QMLP config.}\vspace{-6pt}
    \label{f:noise_impact}
\end{figure}

\textbf{Results on different QMLP configuration}.
Figure~\ref{f:noise_impact} shows the performance of different QMLP configuration. 
First, it shows that models with fixed two-qubit \texttt{CNOT} gates, i.e., \texttt{RX-CNOT}, \texttt{DEEP-RX-CNOT}, and \texttt{RXY-CNOT} only deliver $<$70\% accuracy. In contrast, all QMLP models with parameterized two-qubit \texttt{CRX} gates leads to $>$5\% improvement, showing increased learning capability. 
Second, the comparison between \texttt{RX-CRX} and \texttt{DEEP-RX-CRX} on noise-free simulation shows that RUUs provide an effective way to develop deep QNNs with improved model accuracy. However, as circuits become more complex, performance improvements are offset by the side effect of gate errors when noise is involved. Therefore, the trade-off between model depth and noise tolerance should be carefully considered when designing a practical QMLP model.
In addition, the design of nonlinear RUUs should take into account the influence of noise. For instance, \texttt{RX-CRX} achieved the best performance in noise-free simulation, however, \texttt{RXY-CRXY} preforms the best on noisy devices.

\begin{figure}
    \center{\includegraphics[width=1\linewidth]{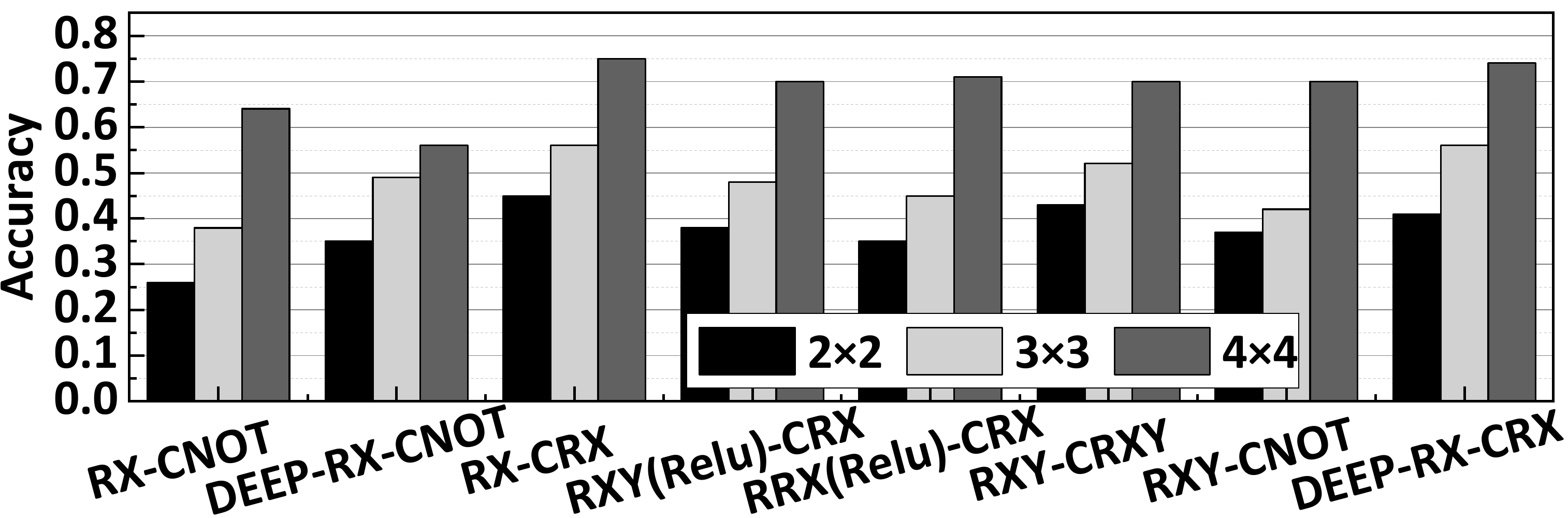}}\vspace{-8pt}
    \caption{The impact of input sizes on different QMLP config.}
    \label{f:scale_input}\vspace{-10pt}
\end{figure}

\textbf{Impact of input sizes}.
We explore the impact of input sized on QMLP performance and show the results in Figure~\ref{f:scale_input}. We down-sample the original MNIST images to 2$\times$2, 3$\times$3, and 4$\times$4, and evaluate the inference accuracy on different QMLP circuits. QMLP performance increase monotonically with the input sizes, which can be intuitively interpreted as the richer the preserved correlations in the low-level quantum input feature maps, the easier it is to be captured by a QNN model to achieve high performance.


\section{Conclusion}
\label{sec:conclusion}
In this paper, we propose QMLP, an error-tolerant quantum network architecture.
QMLP paves the way for implementing scalable deep quantum neural networks with
error-tolerant linear input embedding,
adjustable nonlinearity,
and enhanced variational circuit ansatz.
We evaluate QMLP on MNIST 10-class inference against state-of-the-art quantum neural networks. QMLP achieves 10\% accuracy improvement with 2$\times$ fewer quantum gates and 3$\times$ reduced parameters.


\vspace{-4pt}
\bibliographystyle{ieeetr}
\bibliography{main}

\end{document}